\documentclass[12pt]{iopart}
\usepackage{graphicx}
\begin{document}

\title[]{$\phi$ meson production in $\sqrt{s_{NN}}$ = 200 GeV Au+Au
and pp collisions at RHIC}

\author{Jingguo Ma \dag for the STAR Collaboration   
\footnote[1]{Jingguo Ma (jgma@physics.ucla.edu)}
\footnote[2]{See Ref~\cite{ref0} for full author list}
}

\address{\dag\  Department of Physics and Astronomy, University of
California, Los Angeles, 405 Hilgard Avenue, Los Angeles, CA 90095, 
U.S.A}

\begin{abstract}
We present the results for the measurement of $\phi$ meson  production
in $\sqrt{s_{NN}}$ = 200 GeV Au+Au and pp collisions at the
Relativistic Heavy Ion Collider (RHIC). Using the event mixing 
technique, spectra and yields are obtained from the
$\phi\rightarrow K^{+}K^{-}$ decay channel for different centrality bins in
Au+Au collisions and in pp collisions. We observe that the spectrum 
shape in Au+Au collisions depends weakly on the centrality and the shape
of the spectrum in pp collisions is significantly different from that
in Au+Au collisions. In Au+Au collisions, the extracted yield of
$\phi$ meson is flat as a function of rapidity; The $<p_T>$ of $\phi$, 
extracted from the fit function to the spectra, shows a different
behavior as a function of centrality than that of $\pi^-$, $K^-$ and $\bar{p}$.
\end{abstract}

\submitto{\JPG}
\maketitle

\section{Introduction}
Strangeness production in Relativistic Heavy Ion Collisions may
provide detailed information on the collision dynamics \cite{bib1}. 
The enhanced production of strangeness in nucleus-nucleus collisions
is predicted to be a signature that the collisions go through a
deconfined stage - namely the quark gluon plasma (QGP)
\cite{bib2,bib3}. The $s\bar{s}$ valence quark content for the $\phi$
meson is of particular interest: The production of $\phi$ meson may be 
sensitive to strangeness enhancement \cite{bib2,bib3,bib4}. The
possible mass and decay width modification of $\phi$ meson due to the 
expected partial chiral symetry restoration and in medium effect in
the hot and dense matter has been a topic of both theoretical and 
experimental investigations \cite{bib5,bib6,bib7,bib8,bib9}. The
$\phi$ meson may retain the information from the early partonic stage
of the collisions since it is expected that $\phi$ meson interacts
weakly with nonstrange hadrons during the hadronic stage
\cite{bib3,bib4}. Finally, the comparison of the production of $\phi$
meson in Au+Au and pp collisions at the same beam energy may yield
knowledge on the production mechanisms of $\phi$ meson and the
evolution dynamics of the colliding system. 

\section{Experiment and Analysis}
The results on $\phi$ meson production at $\sqrt{s_{NN}}$ = 130
GeV from the STAR detector have been reported in \cite{bib10}. The 
data presented here was taken by the STAR detector \cite{bib11} 
during the second run of RHIC collider in the year 2001. The main
components of the STAR detector used in this analysis are a large
acceptance Time Projection Chamber (TPC) \cite{bib12}, a Central
Trigger Barrel (CTB) and two Zero Degree Calorimeters. The TPC is 
placed in an uniform magnetic field as the tracking device for
charged particles. The CTB and ZDC are used for triggering. In this
analysis about 2.1M events from minimum bias trigger data and 1.0M
events from central trigger data in Au+Au collisions and 4M events
from minimum bias trigger data in pp collisions are used after all
event selection cuts \cite{bib10}.

  The particle identification is achieved by correlating the measured
energy loss ($dE/dx$) of the particle in the TPC gas with its
momentum. In this analysis, a track is selected as a kaon candidate as
long as its $dE/dx$ is within $2\sigma$ of the kaon Bethe-Bloch
curve. Due to limited resolusion of the detector, the pion and kaon 
dE/dx bands merge at momentum greater than 0.7 GeV, resulting in the
pion contamination in the kaon candidates. The centrality in Au+Au
collisions is defined by the fraction of the total inelastic hadronic 
cross-section, i.e. by dividing the raw charged hadron multiplicity
distribution into centrality classes.

  The $\phi$ signal is built by calculating the invariant mass of
every selected $K^+$$K^-$ pair. The shape of the combinatorial
background is calculated by event mixing technique
\cite{bib13,bib14}. For the Au+Au data analysis, each event is
mixed with another event in the same centrality class, while each
event is mixed with four other events in the pp data analysis in
order to get a better description of the background with higher
statistics.

\section{Results and Discussion}
\vspace{0.0cm}
\begin{figure}[tbh]
\centering\mbox{
\includegraphics[width=0.7\textwidth]{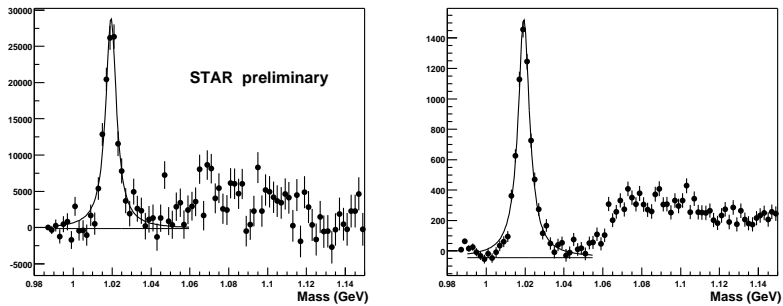}}
\vspace{0.0cm}
\caption{Background subtracted $K^+$$K^-$ pair invariant mass
distribution from the top 10\% central Au+Au collisions (left pannel)
and pp collisions (right pannel). The $\phi$ peak is fitted to a 
Breit-Wigner function plus a linear background function representing
the background. The residual background at higher mass region comes
from misidentified $K_s^0$ decay products.}
\label{invariantmasspeak}
\end{figure}

The background subtracted $K^+$$K^-$ pair invariant mass distribution
averaged over 0.4 GeV/c $\leq p_T \leq$ 3.9 GeV/c is shown in Figure
\ref{invariantmasspeak}. The $\phi$ peak is fit to a Breit-Wigner function plus a
linear function representing the background in each $p_T$ bin. The
measured mass for the  $\phi$ meson in both Au+Au and pp collisions 
is $1019\pm0.7$ MeV/$c^2$, which is consistent with the $\phi$ meson mass 
from the Particle Data Group \cite{bib15}. The measured width is also consistent 
with the $\phi$ natural width convoluted with the resolution of the 
STAR detector. The residual background at mass region greater than
1.06 GeV/$c^2$ mainly comes from $K_s^0$ decay products when pions are 
misidentified as kaons. A small fraction of it may also come from 
$\Lambda$ decays where both of its decay daughters are misidentified
as kaons.

\vspace{0.0cm}
\begin{figure}[tbh]
\centering\mbox{
\includegraphics[width=0.5\textwidth]{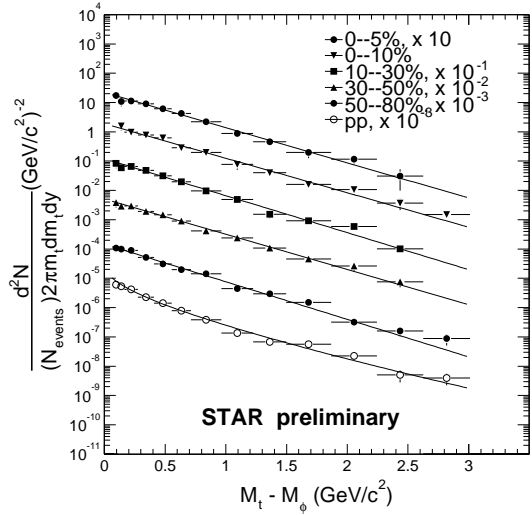}}
\vspace{0.0cm}
\caption{$\phi$ invariant multiplicity distribution as a function of
transverse mass for five centrality bins in Au+Au collisions (filled
symbols) and pp collisions (open symbols) at $\sqrt{s_{NN}}$ = 200
GeV. The spectra are scaled by different factors to guide eyes.}
\label{spectra}
\end{figure}

  In order to obtain the resonance yield, detector acceptance and
efficiency corrections were applied to the uncorrected number of
$\phi$ for each centrality and $p_T$ bin. The acceptance and efficiency
corrections were done by embedding simulated kaons from $\phi$ decays
into real events using GEANT, and by passing them through the full
reconstruction chain \cite{bib10,bib16}. The corrected $\phi$ invariant
multiplicity distributions at mid-rapidity ($|y| <$ 0.5) as a function
of $m_T - m_{\phi}$ are depicted in Figure \ref{spectra}, where
$m_{\phi}$ = 1019.4 MeV/c$^2$ is the average $\phi$ mass reported in
\cite{bib15}. The transverse momentum coverage for this measurement is
0.4 $\leq p_T \leq$ 3.9 GeV/$c$, which corresponds to 85$\%$ of the
$\phi$ yield at mid-rapidity. In Au+Au collisions, the spectra are
fitted by an exponential function in $m_T - m_{\phi}$ for all
centrality bins. In pp collisions, the spectrum is better represented
by a power law function in $p_T$. Thus the spectra shape changes from
pp to Au+Au collisions. In Au+Au collisions, the inverse slopes
extracted from the exponential function fit to the spectra do not
depend strongly on the collision centrality.

\vspace{0.0cm}
\begin{figure}[tbh]
\centering\mbox{
\includegraphics[width=0.5\textwidth]{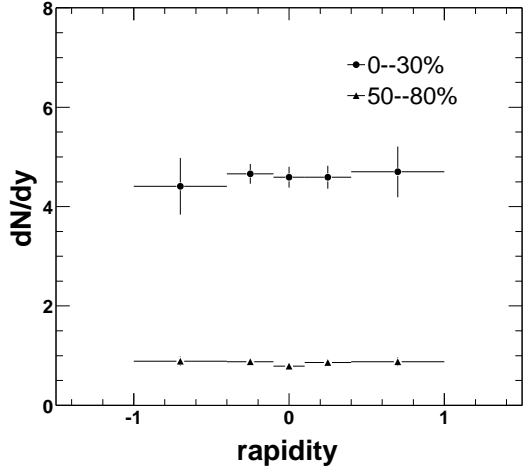}}
\vspace{0.0cm}
\caption{$\phi$ dN/dy as a function of rapidity for Au+Au
collisions. The error bars shown are statistical errors only.
Within the error bars, dN/dy of $\phi$ is flat vs. rapidity for both
0-30\% and 30-80\% centrality bins.}
\label{dndyvsrap}
\end{figure}

  The $\phi$ yield ($dN/dy$) obtained from the fit to the spectra in
Au+Au collisions as a function of rapidity is shown in Figure
\ref{dndyvsrap}. The measured $\phi$ dN/dy is flat as a function of
rapidity within the errors, which is consistent with boost invariance
for the $\phi$ production in Au+Au collisions at RHIC energy in the
measured rapidity range.

\vspace{0.0cm}
\begin{figure}[tbh]
\centering\mbox{
\includegraphics[width=0.5\textwidth]{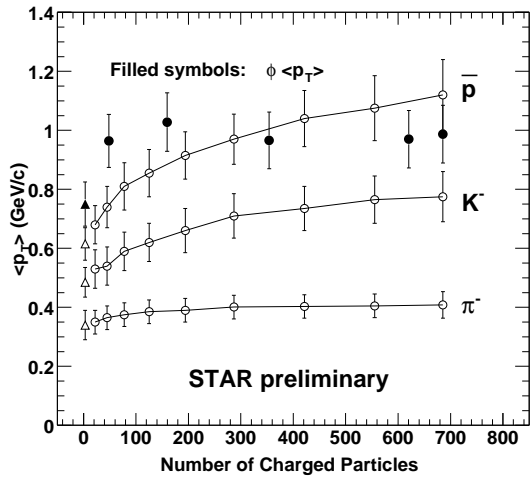}}
\caption{The $<p_T>$ of $\phi$, $\pi^-$, $K^-$ and $\bar{p}$ at
$\sqrt{s_{NN}}$ = 200 GeV Au+Au and pp collisions as a function of
the number of charged particles. The number of charged particles
corresponds to the acceptance and tracking efficiency corrected
charged particle multiplicity within $|\eta| <$ 0.5, where $\eta$ is 
the pseudorapidity. The triangles represent the $<p_T>$ 
obtained in pp interactions and the circles correspond to the $<p_T>$
measured in Au+Au collisions for different centralities. $\phi$
$<p_T>$ in pp collisions is extracted from power law $p_T$ fit
function. In Au+Au collisions, $\phi$ $<p_T>$ is extracted from
exponential $m_T$ fit function. $\phi$ $<p_T>$ shows a very different
behaviour when compared to $\pi^-$, $K^-$ and $\bar{p}$.}
\label{meanpt}
\end{figure}

  The $\phi$ $<p_T>$ calculated from the fits to the spectra as a
function of the number of charged particles is depicted in Figure
\ref{meanpt}, where the solid triangle and the solid circles
sketch the measurements in pp and Au+Au collisions, respectively. In
this figure, the $<p_T>$ of $\pi^-$, K$^-$ and $\bar{p}$ from
\cite{bib17} are also plotted, where the open triangles represent the 
$<p_T>$ obtained in pp interactions and the open circles correspond to
the $<p_T>$ measured in Au+Au collisions for different centralities. 
The $<p_T>$ of $\pi^-$, K$^-$ and $\bar{p}$ increases from peripheral 
to central collisions and from lighter to heavier particles. The 
$\pi^-$ $<p_T>$ increase from pp and most peripheral Au+Au collisions
to the most central Au+Au collisions is about 10$\%$. In the case of 
K$^-$ and $\bar{p}$, the increase is about 25$\%$ and 60$\%$, 
respectively. This behavior is expected from collective flow picture 
with hadronic rescattering. The $\phi$ meson has a very different 
behavior: Its $<p_T>$ increases from pp collisions to Au+Au
collisions. However, there is little increase from peripherial Au+Au
to central Au+Au collisions although its mass is even higher than that 
of $\bar{p}$. Some hadronic hydrodynamic calculations, for example
\cite{bib18} predicts collective flow expansion in proportion to
particle mass, such as those exhibited by $\pi^-$, $K^-$ and
$\bar{p}$. The centrality dependence of $<p_T>$ for the $\phi$ meson
shows a distintive difference from that for the $\pi^-$, $K^-$ and
$\bar{p}$. This may indicate that the evolution dynamics of the $\phi$
meson is less sensitive to hadronic rescattering.


\section{Summary}
We have presented rusults on $\phi$ meson production at mid-rapidity 
in Au+Au and pp collisions at $\sqrt{s_{NN}}$ = 200 GeV. The
spectrum in pp collisions is significantly different from that in
Au+Au collisions. The spectra in Au+Au collisions weakly depend
on the collision centralities. The extracted $\phi$ yield is flat as a
function of rapidity, which is consistent with boost invariance for
the $\phi$ production in the measured rapidity range. The weak
dependence of $\phi$ $<p_T>$ on the centralities in Au+Au collisions
is consistent with the expectation that $\phi$ does not interact
strongly with nonstrange hadronic matter.
 
\ack
We wish to thank the RHIC Operations Group and the RHIC Computing Facility
at Brookhaven National Laboratory, and the National Energy Research 
Scientific Computing Center at Lawrence Berkeley National Laboratory
for their support. This work was supported by the Division of Nuclear 
Physics and the Division of High Energy Physics of the Office of Science of 
the U.S. Department of Energy, the United States National Science Foundation,
the Bundesministerium fuer Bildung und Forschung of Germany,
the Institut National de la Physique Nucleaire et de la Physique 
des Particules of France, the United Kingdom Engineering and Physical 
Sciences Research Council, Fundacao de Amparo a Pesquisa do Estado de Sao 
Paulo, Brazil, the Russian Ministry of Science and Technology, the
Ministry of Education of China, the National Natural Science Foundation 
of China, and the Swiss National Science Foundation. This work has
been partially supported by NSF grant PHY-03-11859.

\end{document}